\documentclass[iop,apj,tighten]{emulateapj}
\usepackage{natbib}
\usepackage{amsmath,amstext}
\usepackage{graphicx} 
\usepackage{comment}

\usepackage{amssymb}

\begin{document}
\title[CN-strong Stars]{Chemical and kinematic analysis of CN-strong Metal-poor Field Stars \\in LAMOST DR3  
}
\author{Baitian Tang\altaffilmark{1}}
\author{Chao Liu\altaffilmark{2}}
\author{J. G. Fern\'andez-Trincado\altaffilmark{3,4,11}}
\author{Doug Geisler\altaffilmark{3,5,6}}
\author{Jianrong Shi\altaffilmark{2}}
\author{Olga Zamora\altaffilmark{7,8}}
\author{Guy Worthey\altaffilmark{9}}
\author{Edmundo Moreno\altaffilmark{10}}

\affil{$^1$School of Physics and Astronomy, Sun Yat-sen University, Zhuhai 519082, China; tangbt@mail.sysu.edu.cn}
\affil{$^2$Key Lab of Optical Astronomy, National Astronomical Observatories, Chinese Academy of Sciences, Beijing 100012, China}
\affil{$^3$Departamento de Astronom\'{i}a, Casilla 160-C, Universidad de Concepci\'{o}n, Concepci\'{o}n, Chile}
\affil{$^{4}$Institut Utinam, CNRS UMR 6213, Universit\'e Bourgogne-Franche-Comt\'e, OSU THETA Franche-Comt\'e, Observatoire de Besan\c{c}on, BP 1615, 25010 Besan\c{c}on Cedex, France}
\affil{$^{5}$Instituto de Investigaci\'{o}n Multidisciplinario en Ciencia y Tecnología, Universidad de La Serena. Avenida Ra\'{u}l Bitr\'{a}n S/N, La Serena, Chile}
\affil{$^{6}$Departamento de F\'{i}sica y Astronom\'{i}a, Facultad de Ciencias, Universidad de La Serena. Av. Juan Cisternas 1200, La Serena, Chile}
\affil{$^7$Instituto de Astrof\'isica de Canarias, 38205 La Laguna, Tenerife, Spain}
\affil{$^8$Departamento de Astrof\'isica, Universidad de La Laguna, 38206 La Laguna, Tenerife, Spain}
\affil{$^{9}$Department of Physics and Astronomy, Washington State University, Pullman, WA 99163-2814, USA}
\affil{$^{10}$Instituto de Astronom\'{i}a, Universidad Nacional Aut\'{o}noma de M\'{e}xico, Apdo. Postal 70264, M\'{e}xico D.F., 04510, M\'{e}xico}
\affil{$^{11}$Instituto de Astronom\'ia y Ciencias Planetarias, Universidad de Atacama, Copayapu 485, Copiap\'o, Chile.}





\begin{abstract}
The large amount of chemical and kinematic information available in large spectroscopic surveys have inspired the search for chemically peculiar stars in the field. Though these metal-poor field stars ([Fe/H$]<-1$) are commonly enriched in nitrogen, their detailed spatial, kinematic, and chemical distributions suggest that various groups may exist, and thus their origin is still a mystery. To study these stars statistically, we increase the sample size by identifying new CN-strong stars with LAMOST DR3 for the first time. We use CN-CH bands around 4000 \AA~to find CN-strong stars, and further separate them into CH-normal stars (44) and CH-strong (or CH) stars (35). The chemical abundances from our data-driven software and APOGEE DR 14 suggest that most CH-normal stars are N-rich, and it cannot be explained by only internal mixing process. The kinematics of our CH-normal stars indicate a substantial fraction of these stars are retrograding, pointing to an extragalactic origin. The chemistry and kinematics of CH-normal stars imply that they may be GC-dissolved stars, or accreted halo stars, or both. 

\end{abstract}

\maketitle

\section{Introduction}
\label{sect:intro}

In recent years, large Galactic surveys with multi-object spectrographs have greatly improved our knowledge about the chemical and kinematic properties of the Milky Way. Sloan Digital Sky Survey (SDSS) is a pioneer and active actor in this field. Beginning from SDSS-II/SEGUE \citep{Yanny2009} and SDSS-III/SEGUE-2, SDSS observed stars in our Milky Way with low resolution optical spectra ($\lambda=3850-9200$ \AA, $R\sim 2000$). Later, the need for higher spectral resolution and the capability to see through the dusty part of our Galaxy have inspired the Apache Point Observatory Galactic Evolution Experiment \citep[APOGEE,][]{Majewski2017} during SDSS-III \citep{Eisenstein2011}. The multi-object NIR fiber spectrograph on the 2.5 m telescope at Apache Point Observatory \citep{Gunn2006} delivers high-resolution ($R\sim$22,500) $H$-band spectra ($\lambda = 1.51 - 1.69$ $\mu$m). The APOGEE program was extended in SDSS-IV as APOGEE-2, which includes observations from northern hemisphere (Apache Point observatory) and southern hemisphere (Las Campanas Observatory).

Searching for stars with peculiar chemistry in the field has recently become tractable, owing to the massive amount of spectra/chemical abundances available in spectroscopic surveys. 
\citet{Martell2010, Martell2011} (hereafter, M11 for these two papers) used CN, CH molecular bands to search for CN-strong stars in SDSS-II/SEGUE and SDSS-III/SEGUE-2. They suggested that these CN-strong stars come from globular clusters (GCs), and a minimum of 17\% of the present-day mass of the stellar halo was originally formed in GCs. Later, \citet{Carollo2013} showed that these CN-strong field stars and the majority of globular clusters exhibit kinematics and orbital properties similar to the inner-halo population.
In fact, most of the scenarios \citep[e.g.,][]{Ventura2011,Ventura2013, Decressin2007, deMink2009, Denissenkov2014} that aim to explain multiple populations (MPs) found in GCs \citep[e.g.,][]{Meszaros2015, Schiavon2017, Tang2017, Tang2018, FT2018} imply a significant mass loss before the enriched stellar generation is formed \citep{Bastian2015a, Renzini2015}. Even if these MP GC scenarios are not considered, due to the tidal force exerted by our Galaxy, GCs are continuously losing stars and some GCs may be disrupted and form observable stellar streams \citep[e.g.,][]{Malhan2018,Ibata2018}. In that sense, it is reasonable to expect to uncover stars now in the field that resemble so-called second generation (SG) stars with enhanced N, Na and depleted C, which presumably are only formed in the dense environments of GCs but then can escape and become field stars.

\begin{figure*}
\centering
\includegraphics [width=0.95\textwidth]{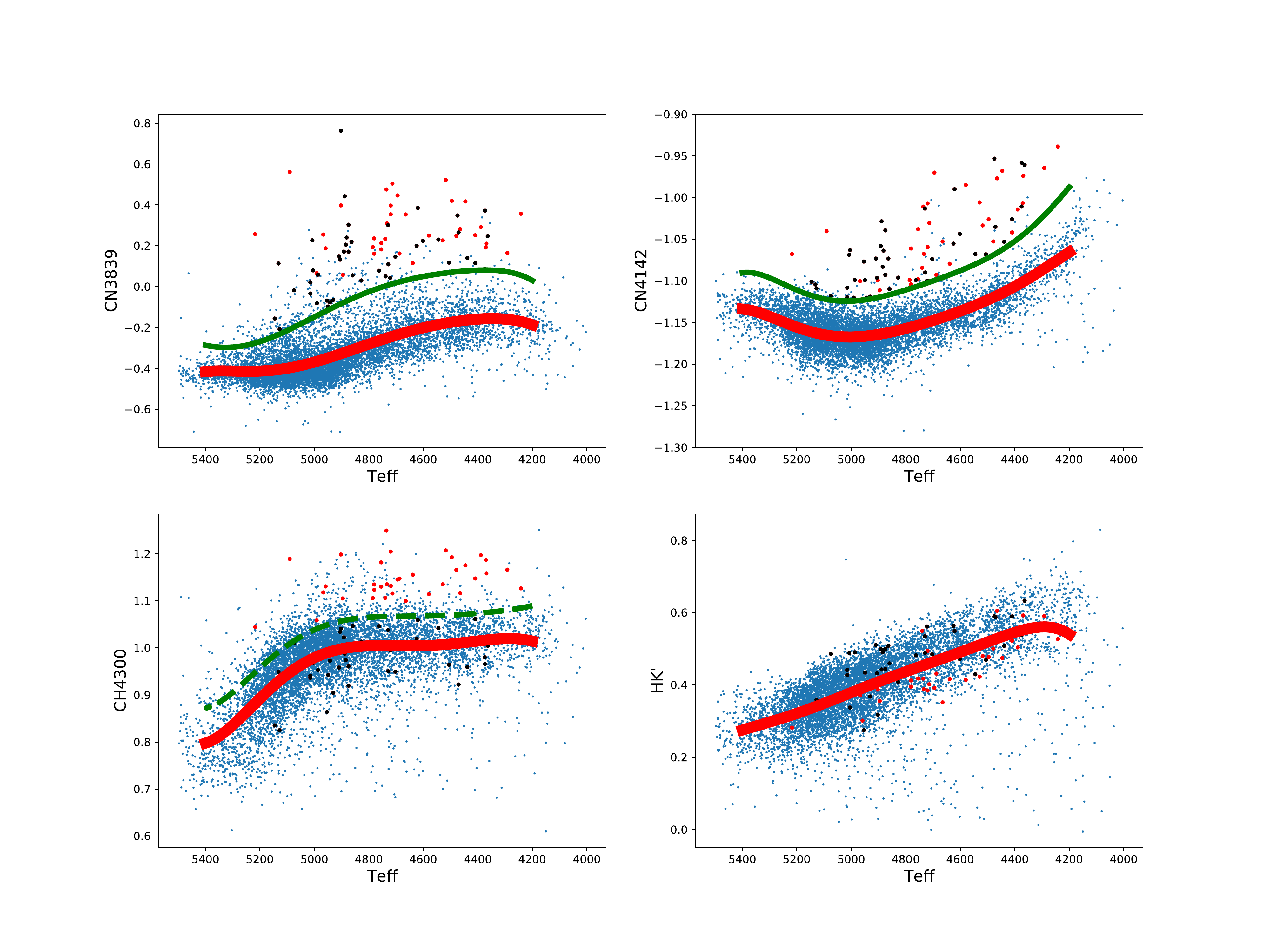} 
\caption{Spectral indices as a function of T$_{\rm eff}$. The metal-poor field stars are shown as blue small dots.
Red lines are sixth-order polynomials of the mean spectral indices at a step of 100K.  Green solid lines are sixth-order polynomials of the mean spectral indices plus 2 times standard deviations at a step of 100K. Similarly, green dashed lines are $\rm mean+1.0\times std$. The CH-strong CN-strong stars are labelled as red dots, and the CH-normal CN-strong stars are black dots. See text for more details.
}\label{fig:CNCH}
\end{figure*}

\citet[][hereafter S17]{Schiavon2017chem} found a large group of N-rich stars in the inner Galaxy with N, C, and Al abundances that are typically found in GC SG stars using APOGEE data. They argued that these stars imply the absence of a mandatory genetic link between SG stars and GCs. Or alternatively, these N-rich stars could be the by-products of chemical enrichment by the first stellar generations.
\citet{Fernandez-Trincado2016, FT2017} further discovered a group of stars with not only high N and Al abundances, but also depleted Mg abundances. These authors separated their sample into two: a metal-rich sample ([Fe/H$]\gtrsim -1.0$) and a metal-poor sample ([Fe/H$]< -1.0$). The stars in the metal-poor sample may find their chemical counterparts in SG Galactic GC stars, but it does not apply to the metal-rich sample. It was speculated that the metal-rich sample stars may (1) migrate from nearby dwarf galaxies, and at the same time contaminated by AGB companion stars; or (2) come from dissolved extragalactic GCs.


While astronomers found an increasing number of N-rich stars using SDSS data, the LAMOST Galactic spectroscopic survey \citep{Zhao2012, Deng2012} has not been explored in this regard.
LAMOST Galactic spectroscopic survey observes stars with low-resolution ($R\sim1800$) optical ($\lambda=3700-9000$ \AA) spectra. Its ability to reach fainter objects and its large sample size make LAMOST survey appropriate for Galactic halo star studies \citep[e.g.,][]{Liu2017}. LAMOST Stellar Parameter pipeline \citep[LASP,][]{Wu2011, Luo2015} is able to determine radial velocity (RV), T$_{\rm eff}$, $\log g$, and [Fe/H]. \citet{Luo2015} compared RVs and stellar parameters for stars in common between LAMOST DR1 and APOGEE. The typical uncertainties for these parameters derived from LAMOST data are 4 km/s for RV,100 K for T$_{\rm eff}$, 0.25 dex for $\log g$, and 0.1 dex for [Fe/H], respectively. The LAMOST data release is now in its fifth phase (DR5), and mid-resolution spectral observations ($R\sim 7500$) have been implemented recently. Here we use the public data release, DR3, which includes the data taken between October 2011 and May 2015. DR3 was updated in June 2017 with a few minor corrections. More specifically, we use the A, F, G and K type star catalog in our study, where more than 3 million stars are included.

In this work, we focus on the CN-strong metal-poor field stars from the LAMOST DR3 (Section \ref{sect:data}).  These CN-strong stars  are further separated into CH-strong and CH-normal stars, because they show different properties. We exam these stars carefully using the chemical abundances derived by the APOGEE team and our data-driven software (Section \ref{sect:exam}). In Section \ref{sect:sp} and \ref{sect:orb}, the origins of CH-normal stars are discussed using their spatial distribution and orbital configuration. We further discuss the origins of these CH-normal stars using additional observational evidence as well in Section \ref{sect:dis}.  A brief summary is given in Section \ref{sect:con}.

\begin{figure*}
\centering
\includegraphics [width=0.95\textwidth]{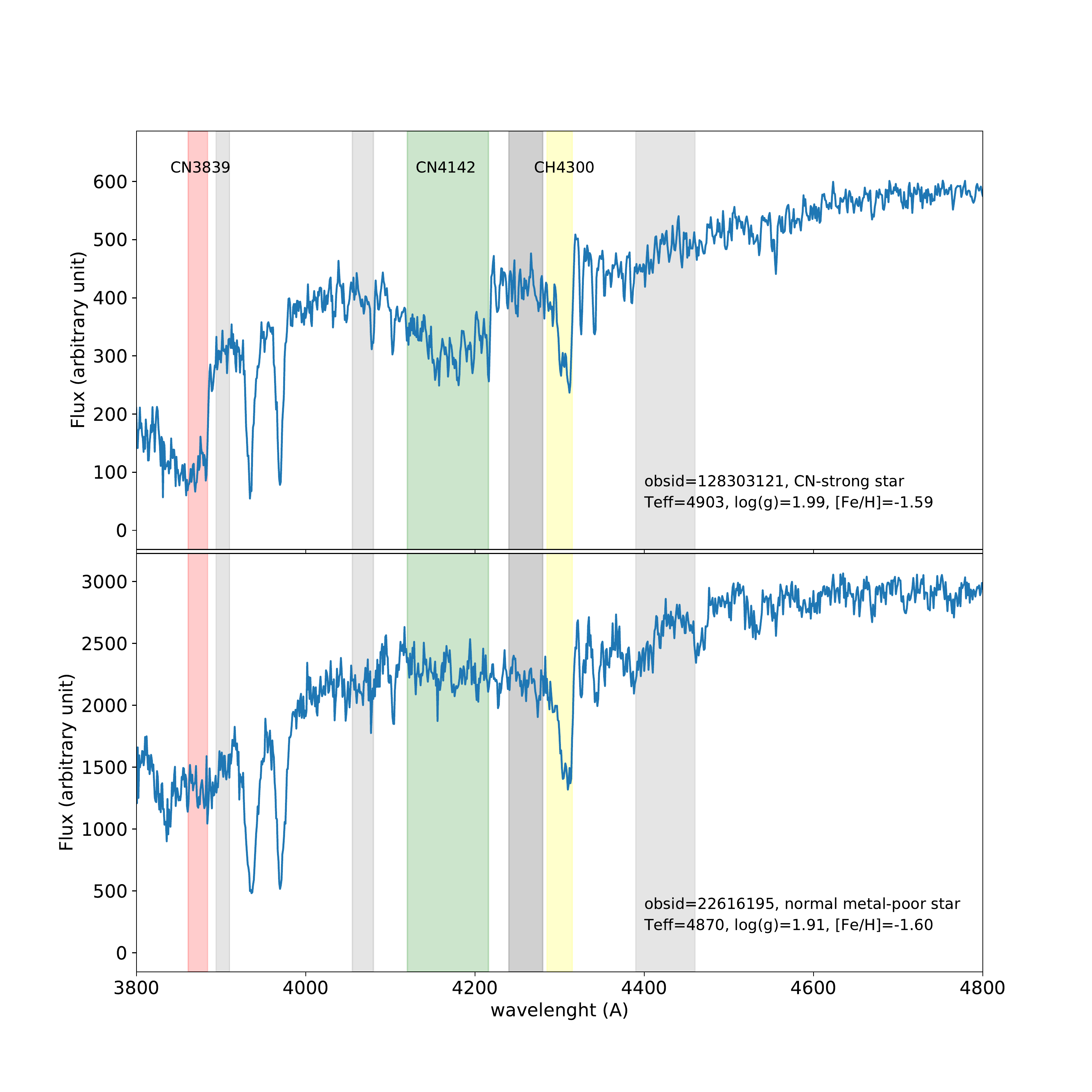} 
\caption{Spectra of one CN-strong star (top panel) and one normal metal-poor star (bottom panel) between 3800 \AA~and 4800 \AA. The feature wavelengths of CN3839, CN4142, CN4309 are labeled with red, green, and yellow colors, respectively. The pseudo-continuum regions are labeled with gray color. The LAMOST observed ID and stellar parameters are given in the low right of each panel.
}\label{fig:spec}
\end{figure*}

\section{Sample Selection and Data Reduction}
\label{sect:data}

\begin{figure*}
\centering
\includegraphics [width=0.85\textwidth]{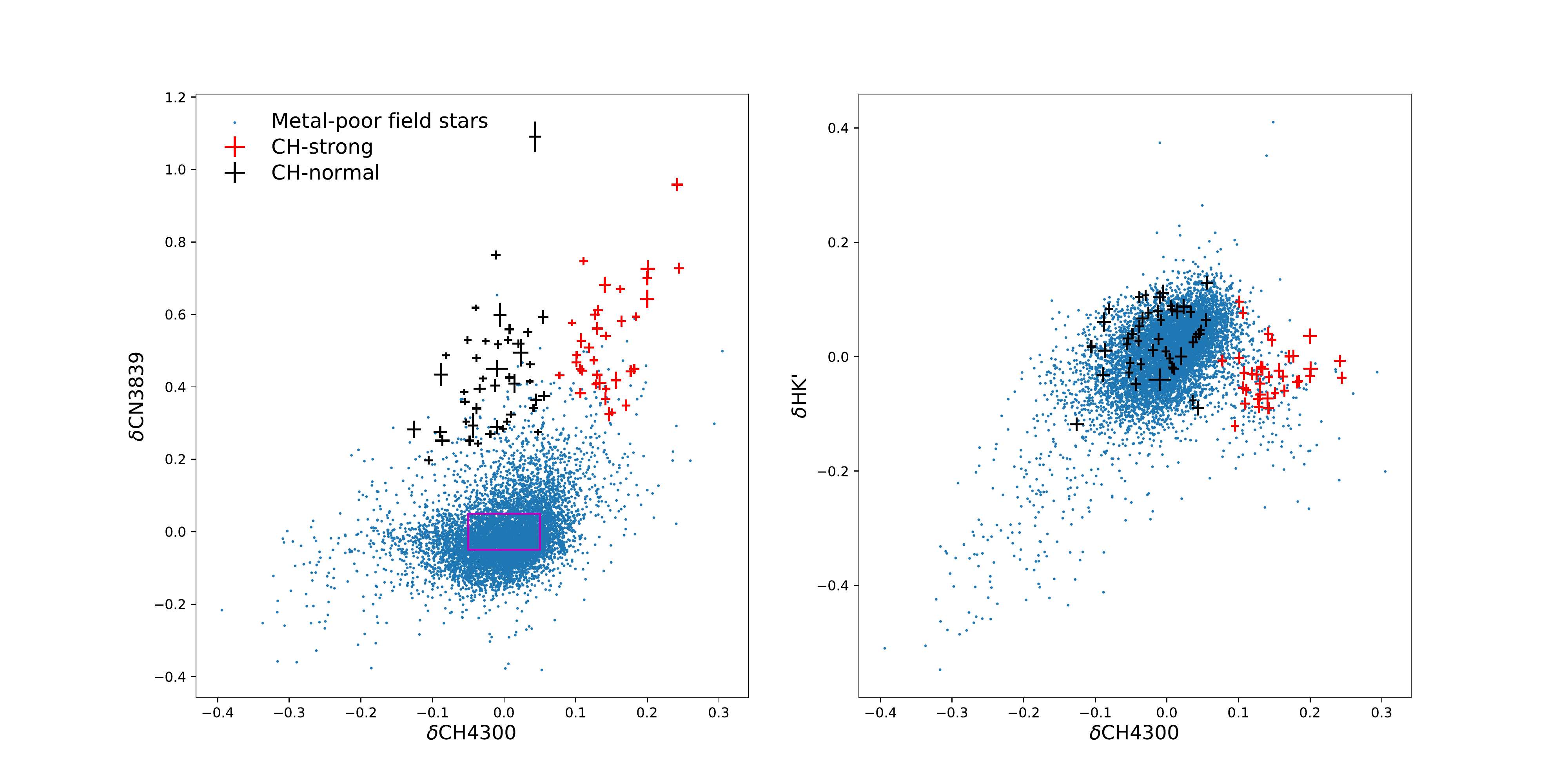} 
\caption{$\delta$CN3839 and $\delta$HK$^{\prime}$ vs. $\delta$CH4300. The CH-strong and CH-normal stars are labelled as red and black symbols, respectively, where the errorbars indicate the measurement uncertainties.  The metal-poor field stars are shown as blue small dots. A sample of normal metal-poor field stars are selected based on their $\delta$CH4300 and $\delta$CN3839 indices detailed by the red box.}\label{fig:dchdcn2}
\end{figure*}

\begin{figure}
\centering
\includegraphics [width=0.45\textwidth]{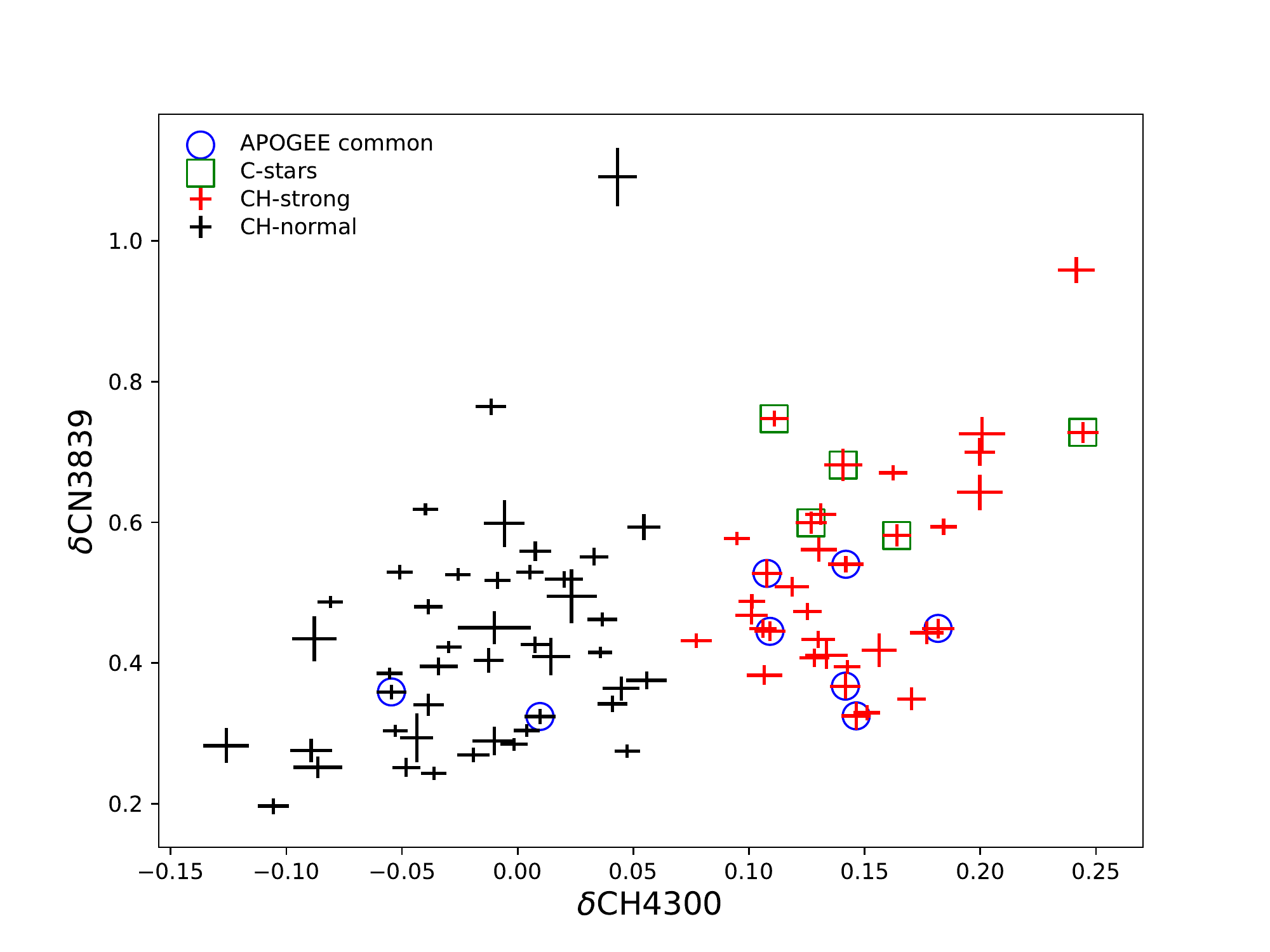} 
\caption{$\delta$CN3839 vs. $\delta$CH4300. The CH-strong and CH-normal stars are labelled as red and black symbols, respectively, where the errorbars indicate the measurement uncertainties. Stars in common with APOGEE DR14 are indicated by blue circles. Stars in common with \citet{LiYB2018} are labelled by green squares.
}\label{fig:dchdcn}
\end{figure}

Given that we are interested in CN-strong metal-poor field stars, we first select metal-poor field stars from the A, F, G and K type star catalog of LAMOST DR3, following these criteria:
\begin{enumerate}
\item $4000<T_{\rm eff}<5500$ K
\item $\log$ g $< 3.0$
\item $-1.8<[$Fe/H$]<-1.0$
\item SNR$_u>5.0$
\end{enumerate} 
The last one indicates that the signal to noise ratio (SNR) at u band is greater than 5.0, which ensures that we have at least mid-quality spectra around 4000 \AA~for spectral analysis. The metallicity range is similar to that of M11. The stellar parameters and SNR$_u$ are provided as part of LAMOST DR3. We further confirm that these stars are not members of known GCs by implementing the GC selection method described in \citet{Tang2017}. Moreover, since the stars that we select are metal-poor ([Fe/H$]<-1.0$), we do not expect them to be open cluster members either. When correcting for RV in the spectra, we also consider the systematic RV shifts ($\sim 5$ km/s) reported for the LAMOST spectra \citep{Schonrich2017}.
Next, we measure the spectral indices of our sample stars. Here we use the definition of CN3839, CN4142, and CH4300 indices from \citet{Harbeck2003} and HK$^{\prime}$ from \citet{Lim2015}. To estimate the error in index measurement, we use the Monte-Carlo (MC) simulation, where flux-associated error is used as standard deviation of the random error in MC. The final spectral index error is the quadratic sum of error caused by flux uncertainty and error caused by RV uncertainty.

\begin{figure*}
\centering
\includegraphics [width=0.95\textwidth]{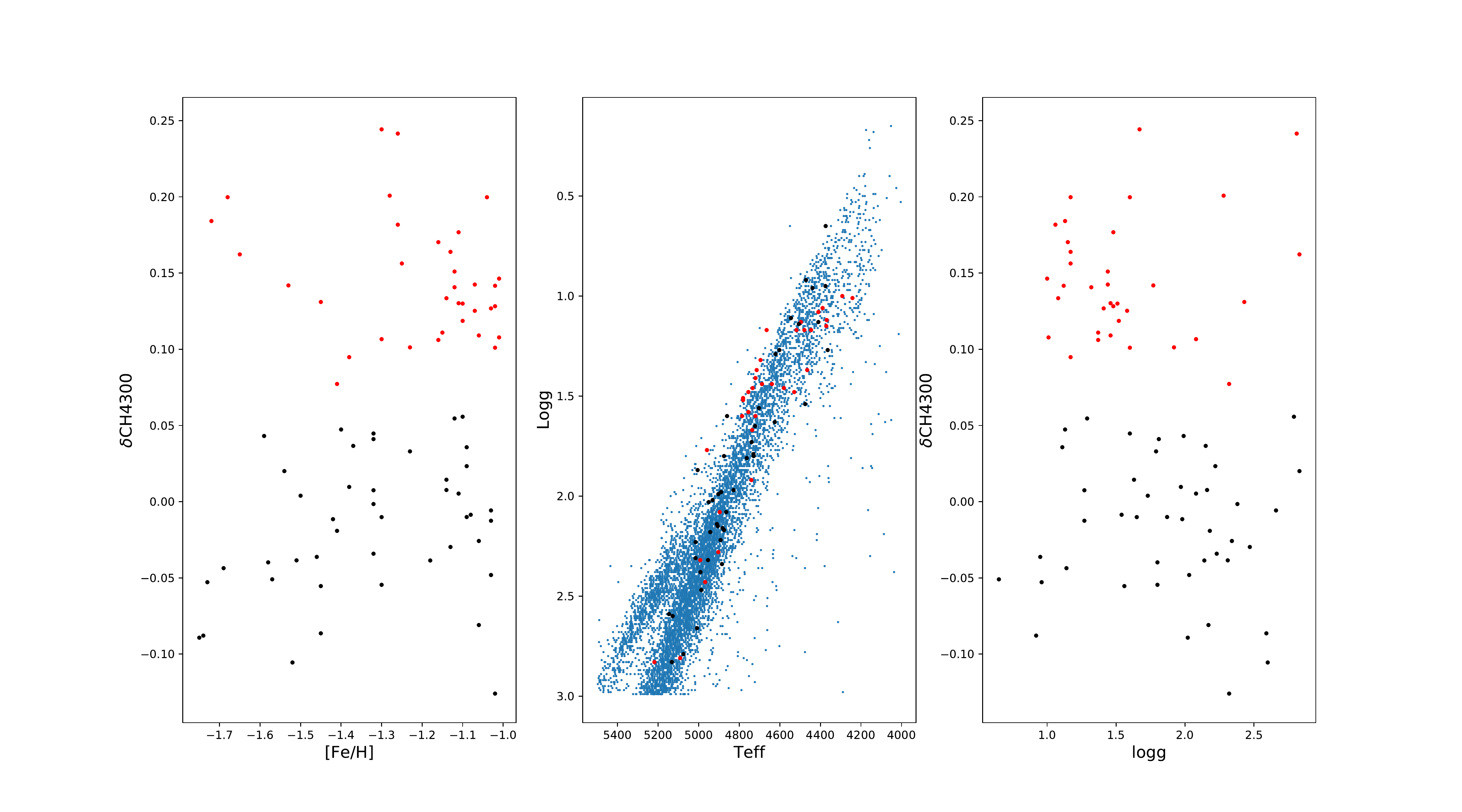} 
\caption{Relations between $\delta$CH4300 and stellar parameters. Symbols have similar meanings as Figure \ref{fig:CNCH}.
}\label{fig:tlogg}
\end{figure*}

Since the spectral indices that we use here are closely related to T$_{\rm eff}$\footnote{ $\log g$ and metallicity also have minor effects. The effect of $\log g$ is weakened when only giants are concerned.}, we plot them as a function of T$_{\rm eff}$ in Figure \ref{fig:CNCH}. To find out the N-rich halo stars, we calculate the means and standard deviations (std for short) for CN3839 and CN4142 indices, at a step of 100K, and fit the mean and $\rm mean+2.0\times std$ as a function of T$_{\rm eff}$ with sixth-order polynomials (Figure \ref{fig:CNCH} red and green solid lines, respectively). We select stars above the green solid line for both CN3829 and CN4142 indices as our final sample of CN-strong stars. Seventy-nine stars are selected. We use both CN indices here to exclude stars whose spectra may be influenced by strange emission lines or other unwanted features around the CN bands. We also exam all 79 stars one-by-one to make sure that the absorption bands are reliable. For example, Figure \ref{fig:spec} compares one CN-strong star (top panel) with one normal metal-poor field star (bottom panel). These two stars are picked with similar stellar parameters to minimize the effects of different stellar parameters on the spectra. The stark difference between these two stars near the feature wavelengths of CN3839 and CN4142 confirms that our classification is based on real spectral signals.

Among these CN-strong stars, there are CH-strong and CH-normal stars. We determine the mean and std for CH4300, at a step of 100K, and fit the mean and $\rm mean+1.0\times std$ with sixth-order polynomials (Figure \ref{fig:CNCH} red solid line and green dashed line, respectively). 
We label the CH-strong stars with red dots, and the CH-normal stars with black dots\footnote{Unless noted otherwise, we will refer CN-strong CH-strong stars as CH-strong stars, and CN-strong CH-normal stars as CH-normal stars in this paper.}. Later in this work, we show that these two groups of stars have other distinct properties.  We note that distinguishing CH-normal stars from CH-strong stars has been implemented in other similar studies. For example, M11 excluded stars with strong CH feature around 4350 \AA~and strong C$_2$ band at 4737 \AA; S17 and \citet{FT2017} excluded stars with [C/Fe$]>+0.15$ dex.
Next, we define $\delta$CH4300 as the CH4300 index value minus the mean polynomial fit value at the T$_{\rm eff}$ of a given star. A similar definition is also applied to CN3839 and HK$^{\prime}$. We use CN3839 in this work, because CN3839 is suggested to show higher sensitivity and smaller error than CN4142 \citep{Harbeck2003,Pancino2010}.

We notice that CH-strong stars tend to have lower $\delta$HK$^{\prime}$ (Figure \ref{fig:dchdcn2}).  Comparing with the bulk of metal-poor field stars, these CH-strong stars occupy different location in the $\delta$HK$^{\prime}-\delta$CH4300 parameter space, while the CH-normal stars follow the trend defined by most of the metal-poor field stars.  The HK$^{\prime}$ index is carefully discussed in \citet{Lim2015}, where they showed that this index measures only calcium lines, and the contamination from CN band is minimized. Therefore, we do not expect $\delta$HK$^{\prime}$ bias towards any of these three kinds of stars. 
Furthermore, \citet{LiYB2018} recently identified 2651 carbon stars from more than 7 million spectra in LAMOST DR4. We match their results with our CN-strong star sample, and find five stars in common (green squares in Figure \ref{fig:dchdcn}). Three are labeled as CH stars, while the other two are labeled as barium star in their work. These five carbon stars are all CH-strong stars according to our definition.

Therefore, we consider our method for distinguishing CH-strong stars from CH-normal stars to be efficient. CH-strong stars are different species of stars compared to CH-normal stars and most metal-poor field stars.

\section{Careful Examination of the CN-strong Metal-poor Field Stars}
\label{sect:exam}

\begin{figure*}
\centering
\includegraphics [width=0.85\textwidth]{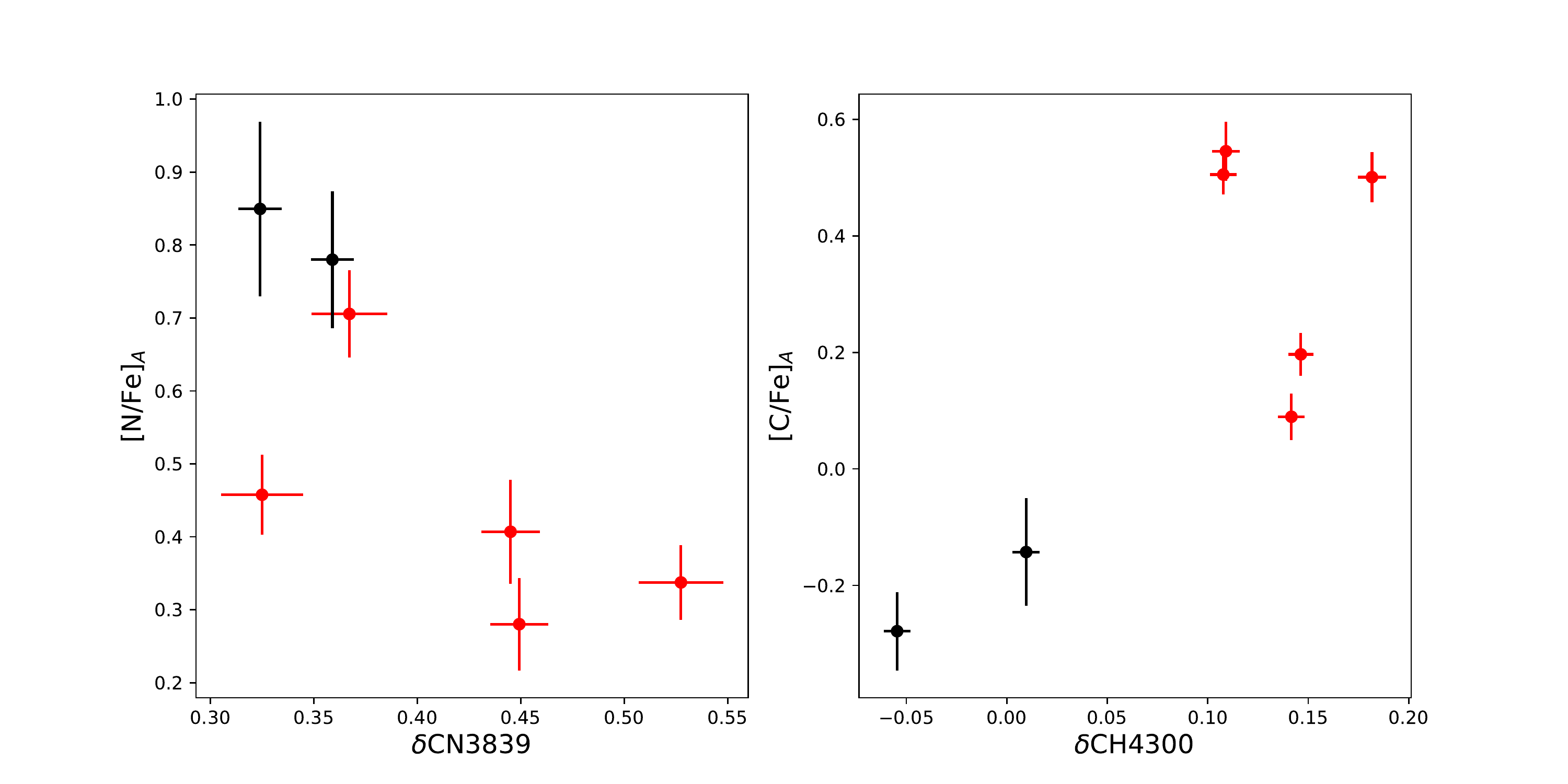} 
\caption{Relations between spectral indices and APOGEE-derived C, N abundances. The CH-strong and CH-normal stars in common with APOGEE DR14 are labelled as red and black symbols, respectively, where the errorbars indicate the measurement uncertainties. 
}\label{fig:comp}
\end{figure*} 

\begin{figure*}
\centering
\includegraphics [width=1.05\textwidth]{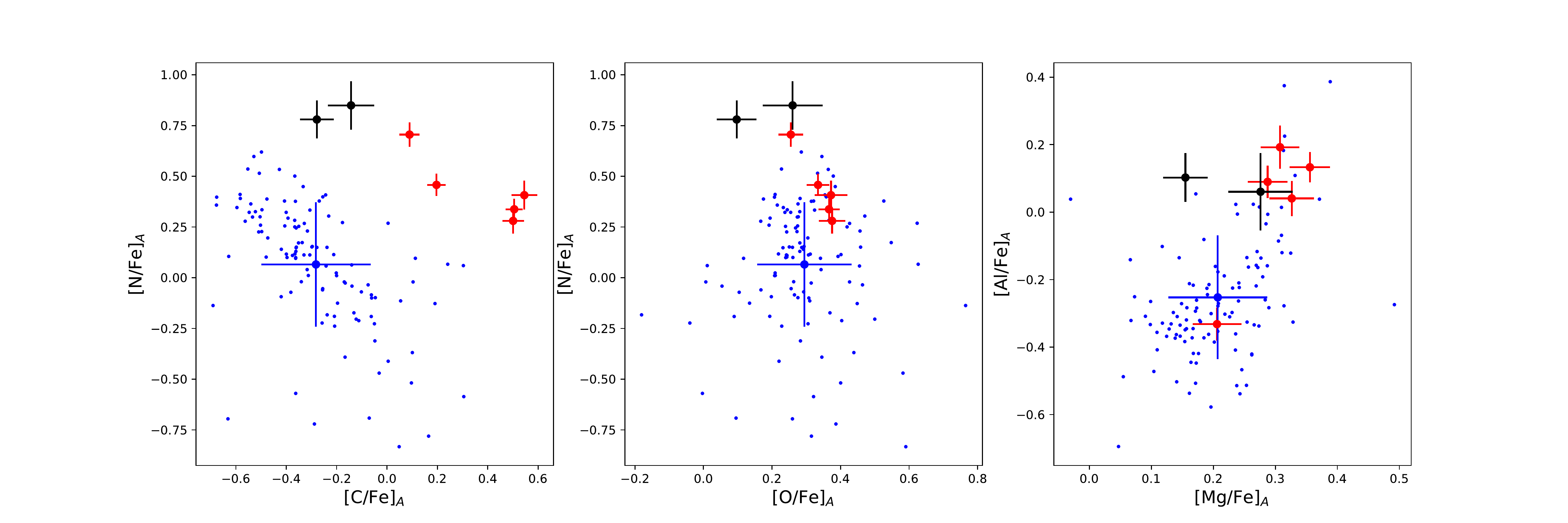} 
\caption{APOGEE-derived chemical abundances. The CH-strong and CH-normal stars in common with APOGEE DR14 are labelled as red and black symbols, respectively, where the errorbars indicate the measurement uncertainties. The normal metal-poor field stars in common with APOGEE DR14 are labelled as blue dots. The blue errorbars indicate their mean and standard deviations of the chemical abundances. 
}\label{fig:abun}
\end{figure*} 

\begin{figure}
\centering
\includegraphics [width=0.45\textwidth]{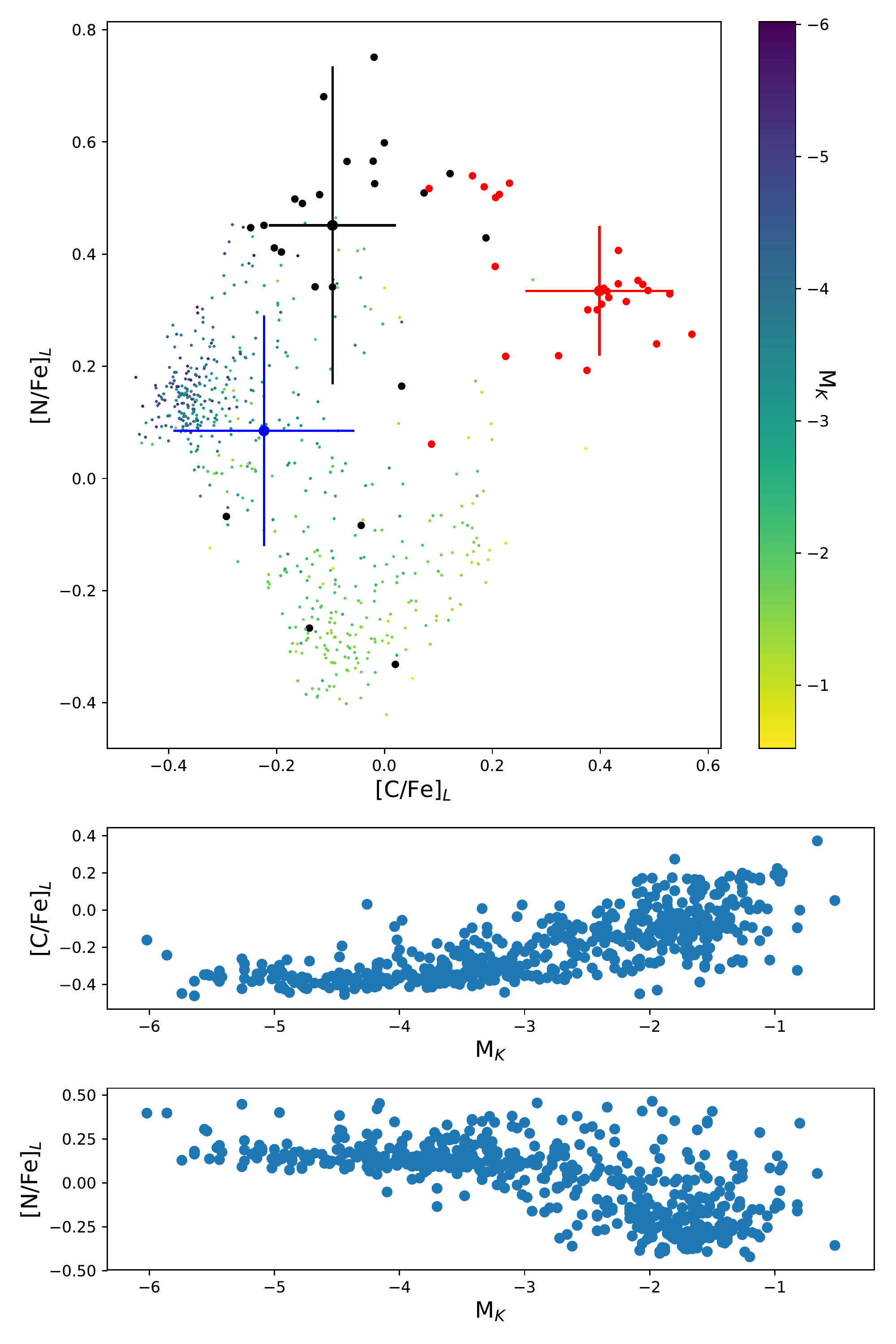} 
\caption{[N/Fe] vs. [C/Fe] using our data-driven software based on LAMOST DR3 data. Subscript `L' denotes `LAMOST'. $Top$ Panel: CH-strong (CH-normal) stars are indicated by black (red) dots. Normal metal-poor field stars are labeled as small dots, where their $K$-band absolute magnitudes are indicated by the color map. It is clear that brighter stars have higher N abundances. The errorbars indicate the means and standard deviations of the chemical abundances of different groups of stars. $Middle$ and $Bottom$ Panels: [C/Fe] and [N/Fe] as a function of $K$-band absolute magnitude for normal metal-poor field stars only.
}\label{fig:lcn}
\end{figure} 

\subsection{Stellar Parameters}
Metallicity plays an important role in determining the properties of a star. Do CH-strong and CH-normal stars have a different metallicity distribution? 
The left panel of Figure \ref{fig:tlogg} suggests that two kinds of stars do not show obvious different metallicity distribution. Next, we examine the T$_{\rm eff}-\log g$ diagram (the middle panel of Figure \ref{fig:tlogg}). Our CN-strong stars have T$_{\rm eff}$ and $\log g$ consistent with other red giants. Note that our CN-strong stars have a median SNR of 120 in $r$-band, where the uncertainties in $\log g$ determination can be reduced to better than $\sim 0.1$ dex \citep{Xiang2017}.  
Finally, we examine their distributions in $\log g$.  We notice that stars with lower $\log g$ tend to have slightly higher $\delta$CH4300, though the correlation between them is weak --- a Spearman rank correlation coefficient of $-0.26$. 

\begin{table*}
\caption{Common Stars between Our CN-strong Star Sample and APOGEE DR14.}              
\label{tab:comm}      
\centering                                      
\begin{tabular}{c c c c c c c c c c c}         
\hline\hline  
\#& APOGEE\_ID & Teff & [C/Fe]& [N/Fe]& [O/Fe]& [Mg/Fe]& [Al/Fe]& [Si/Fe]& [Ca/Fe]& Note\\
\hline 
1 & 2M13233152+4931144 & 4847.0 & -0.28 & 0.78 & 0.10 & 0.15 & 0.10 & 0.17 & 0.12 & CH-normal \\
2 & 2M15590393+4139542 & 4399.2 & 0.50 & 0.28 & 0.38 & 0.33 & 0.04 & 0.40 & 0.26 & CH-strong \\
3 & 2M07330841+3837042 & 4254.3 & 0.51 & 0.34 & 0.37 & 0.36 & 0.13 & 0.33 & 0.25 & CH-strong \\
4 & 2M11394345+2708552 & 5193.4 & 0.78 & 0.71 & -0.14 & 0.28 & -0.30 & 0.38 & 0.20 & CH-strong; too hot \\
5 & 2M15113274+3059083 & 4716.8 & 0.55 & 0.41 & 0.37 & 0.31 & 0.19 & 0.31 & 0.26 & CH-strong \\
6 & 2M13413240-0116003 & 4344.9 & 0.20 & 0.46 & 0.33 & 0.29 & 0.09 & 0.39 & 0.19 & CH-strong \\
7 & 2M12561260+2804017 & 4858.3 & -0.14 & 0.85 & 0.26 & 0.28 & 0.06 & 0.36 & 0.23 & CH-normal \\
8 & 2M19263774+4434325 & 4253.1 & 0.09 & 0.71 & 0.25 & 0.21 & -0.33 & 0.28 & 0.20 & CH-strong \\
\hline                                             
\end{tabular}

\end{table*}

\subsection{Chemical Abundances from APOGEE Pipeline and LAMOST Data-driven Software}

The high-resolution spectroscopic survey APOGEE can provide up to more than 20 elemental abundances \citep{Holtzman2015}, which may improve our understanding of the CN-strong field stars.
After we match our CN-strong field stars with APOGEE DR14 database, we find eight stars in common (Table \ref{tab:comm}). Next, we take a closer look at the chemical abundances of these common stars given by the APOGEE Stellar Parameter and Chemical Abundances Pipeline (ASPCAP; \citealt{GP2016}). Most of the stars have high SNR APOGEE spectra (SNR$>$90), except Star4, partially because its T$_{\rm eff}$ is greater than 5000 K. After visually inspecting the APOGEE spectra, we find that most of the CO, OH, and CN lines are detectable, and the ASPCAP best fits are reasonable. The exception is again found to be Star4, where CNO measurements are no longer reliable for stars hotter than 5000 K \citep[][]{Souto2016}. We also visually examine the Mg and Al lines in these eight common stars, the ASPCAP fits looks reasonable. Therefore, we use the [C/Fe], [N/Fe], [O/Fe], [Mg/Fe], and [Al/Fe] from ASPCAP to study these common stars, without performing manual analysis. 
To define a control sample that represents the normal metal-poor field stars, we select stars that satisfies: (1) $-0.05<\delta$CN3839$<0.05$, (2) $-0.05<\delta$CH4300$<0.05$ (the red box in Figure \ref{fig:dchdcn2}), and (3) T$_{\rm eff}<5000$ K. In total, 1314 stars are selected. We match this sample with APOGEE DR14, and find 115 stars in common. Because we are using the ASPCAP results of these 115 stars without examining the details of each spectrum, we mainly use their statistical mean and standard deviation values.

Figure \ref{fig:comp} shows positive correlation between $\delta$CH4300 and [C/Fe]$_{\rm A}$\footnote{Subscript `A' denotes `ASPCAP'.}, but the correlation between  $\delta$CN3839 and [N/Fe]$_{\rm A}$ is more complicated. It means that $\delta$CH4300 is a good indicator of the C abundance, but $\delta$CN3839 depends on both C and N abundances. 
For the CH-normal stars, the N abundances are the highest of all (Star1 and Star7). 

We will first compare CH-normal stars with normal metal-poor field stars, and turn to CH-strong stars towards the end of this section.
The left panel of Figure \ref{fig:abun} shows interesting physics in the N-C parameter space. Our normal metal-poor field stars, which are mostly giants, show an anti-correlation between [N/Fe]$_{\rm A}$ and [C/Fe]$_{\rm A}$. It is likely these stars are going through extra mixing along the RGB \citep{Gratton2000}, and thus as C decreases, N increases. This anti-correlation is further discussed later with abundances derived from another approach. The normal metal-poor field stars and CH-normal stars show comparable C abundances, but CH-normal stars show much higher N abundances, which is similar to what is found in GC SG stars \citep{Meszaros2015,Tang2017}. We check the Na lines in the APOGEE spectra, but they are too weak for reliable analysis. Thus we lose the chance to use the classical Na-O diagnostic diagram to identify their GC origin. Looking at [O/Fe]$_{\rm A}$ (middle panel of Figure \ref{fig:abun}), CH-normal stars and normal metal-poor field stars have similar O abundances. 
On the other hand, CH-normal stars and normal metal-poor field stars generally show Mg-enhancement ([Mg/Fe$]_{\rm A}\sim 0.30$), consistent with the $\alpha$-enhancement seen in thick disk or halo stars. Interestingly, CH-normal stars tend to show higher than average [Al/Fe]$_{\rm A}$ compared to normal metal-poor field stars. Compared with Figure 1(a) of \citet{FT2017}, our two CH-normal stars show neither depleted Mg nor high Al abundances as the stars in \citet{FT2017}, but do show similar Mg and Al abundances as the N-rich stars of S17. We will further discuss this in Section \ref{sect:dis}.  
To summarize, the [C/Fe], [N/Fe], [O/Fe], [Mg/Fe], and [Al/Fe] abundances from APOGEE support that CH-normal stars are chemically peculiar compared to normal metal-poor field stars.
We realize that our common sample size for CH-normal stars is small.
High-resolution spectroscopic data (e.g., APOGEE) is necessary to increase our common sample size. Furthermore, optical high-resolution data is also necessary to derive Na and s-process element abundances, which are helpful to constrain their nature, e.g., SG origin or extragalactic origin. 

Recently, data-driven software (e.g., $The~Cannon$) are applied to large data sets to derive chemical abundances \citep{Ness2015, Ho2017, Ting2017}. Another data-driven algorithm of stellar parameterization is SLAM (Zhang et al. in preparation). Similar to The Cannon, SLAM provides a forward model to produce a model spectrum from a set of given stellar parameters and then finds the best-fit stellar parameters for the observed spectrum in terms of chi-squared. Unlike The Cannon, SLAM applies a machine learning algorithm, rather than a polynomial,  to set up the forward model, i.e. produces the flux at each wavelength as a function of effective temperature, surface gravity, metallicity, alpha abundance etc. The advantage of a machine learning-based forward model is that it can handle well the situation of highly non-linear data. As a consequence, SLAM is able to predict stellar parameters not only for late-type giant stars, but also for stars with a large range of effective temperature. In the current version of SLAM, which estimates stellar parameters of the K giant stars used in this work, a support vector regression algorithm is applied for the forward model. Meanwhile, SLAM uses about 9000 common K giant stars between LAMOST DR3 and APOGEE DR13 as the training dataset with the spectra from LAMOST  and the stellar parameters from APOGEE  DR13. 
 The precision of stellar parameter estimation from SLAM reaches $\sigma(\rm T_{eff}) \sim 50$ K, $\sigma(\log g) \sim 0.13$ dex, $\sigma(\rm [M/H]) \sim 0.04$ dex, $\sigma(\rm [\alpha/M]) \sim 0.04$ dex, $\sigma(\rm [C/Fe]) \sim 0.09$ dex, and $\sigma(\rm [N/Fe]) \sim 0.1$ dex at SNR$\sim 40$. At SNR$\sim 100$, the precision increases to $\sigma(\rm T_{eff}) \sim 40$ K, $\sigma(\log g) \sim 0.10$ dex, $\sigma(\rm [M/H]) \sim 0.03$ dex, $\sigma(\rm [\alpha/M]) \sim 0.03$ dex, $\sigma(\rm [C/Fe]) \sim 0.06$ dex, and $\sigma(\rm [N/Fe]) \sim 0.07$ dex.

We derived stellar parameters, [C/M], and [N/M] using SLAM. To avoid bad fits at the parameter space edges, we exclude stars with spectral SNR in g band less than 50, and metallicity less than $-1.4$. The derived C and N abundances are shown in Figure \ref{fig:lcn}. Clearly in the top panel, the CH-strong, CH-normal, and metal-poor field stars are separated, and their relative distribution in the N-C parameter space is similar to the case of APOGEE abundances (left panel of Figure \ref{fig:abun}): (1)  Metal-poor field stars form a sequence in the lower left of the top panel.  As evolved stars ascend the red giant branch, C and N abundances may be changed by first dredge up \citep{Iben1964, Iben1967} and extra mixing \citep{ Gratton2000, Charbonnel2010}. Given that for a typical halo/thick disk star of 1 M$_{\odot}$, the first dredge up occurs around T$_{\rm eff}=5200$ K\citep{Boothroyd1999}, and most of our sample stars have T$_{\rm eff}<5000$ K and $\log g < 2.5$, we infer that most stars have already undergone first dredge up.  On the other hand, the C and N abundances of these stars could be altered by extra-mixing. Stars with brighter $K$-band absolute magnitudes tend to have higher [N/Fe] and lower [C/Fe] (middle and bottom panels of Figure \ref{fig:lcn}), which is consistent with extra-mixing theory and observation \citep{ Gratton2000, Charbonnel2010}; (2) CH-normal stars show enhanced median N abundance, and slightly depleted median C abundance. Clearly, the median N abundance of CH-normal stars are enhanced compared to normal metal-poor field stars with similar C abundances. In other words, the enhanced N abundances in CH-normal stars cannot be explained by only extra-mixing effect. We notice that a few CH-normal stars may have low N abundances, probably due to large uncertainties when a particular type of spectra are scarce in the training set, i.e., high N metal-poor stars. 
The statistical similarity between APOGEE C, N abundances and LAMOST derived C, N abundances further strengthen our statement above. 

Next, we take a look at chemical abundances of the CH-strong stars. We have already known that these star have smaller $\delta$HK$^{\prime}$ compared to CH-normal and halo field stars. Figure \ref{fig:abun} and \ref{fig:lcn} suggest that CH-strong stars show higher C, N, Mg, and Al abundances when compared with normal metal-poor field stars. 
The chemistry of CH-strong stars supports the claim that they have different origin than CH-normal stars and normal metal-poor field stars.

 During post main-sequence stellar evolution phases, carbon and nitrogen abundances may be changed by extra mixing, where N abundance increases and C abundance decreases. However, as shown earlier (Section \ref{sect:exam}), the CH-strong and CH-normal stars do not fall into the parameter space defined by normal metal-poor RGB stars going through extra mixing. Later, the C abundance may be changed during AGB phase through third dredge up, when a large amount of carbon is brought up to the surface. However, AGB stars start to dredge up carbon when they have T$_{\rm eff}<4000$ K and $\log g < 0$, which is not applicable to our sample stars.

How should we understand these carbon-enhanced stars? We should not label  them as carbon-enhanced metal-poor (CEMP) stars, since CEMP stars have an upper limit of [Fe/H$]<-1.8$ as described by \citet{Lucatello2005}.  At [Fe/H$]>-1.8$ and log $g<3.0$, carbon-enhanced stars are mostly classical CH stars. The discovery of CH stars can be traced back to the work of \citet{Keenan1942}. Our definition of CH-strong stars in fact matches the original ones of CH stars. Therefore, we will also call our CH-strong stars as CH stars in the following paragraph and in future works. It has been clearly established that CH stars have binary origin \citep{McClure1984, McClure1990}. The peculiar abundance patterns seen in CH stars, combined with the high binary fraction indicate the accretion of material from an intermediate-mass AGB companion, either through Roche lobe overflow or through efficient stellar winds \citep{Han1995}.

\section{Spatial Distribution of CH-normal  Stars}
\label{sect:sp}

\begin{figure}
\centering
\includegraphics [width=0.45\textwidth]{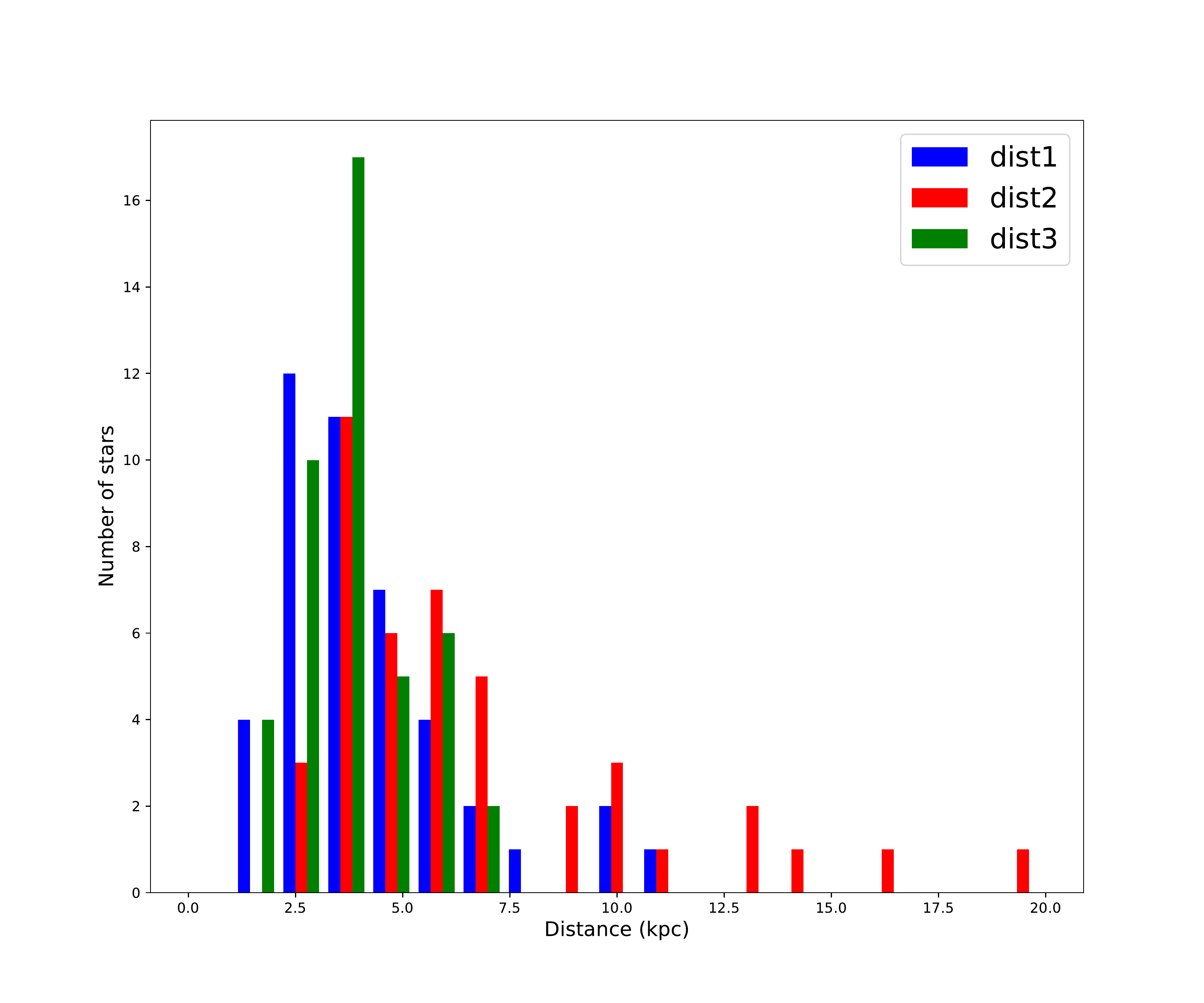} 
\caption{Distance distribution histogram of CH-normal stars using three different methods. Blue, red, and green bars indicate the distance distribution of stars using dist1, dist2, and dist3, respectively (see text). 
}\label{fig:histdist}
\end{figure} 

\begin{figure}
\centering
\includegraphics [width=0.45\textwidth]{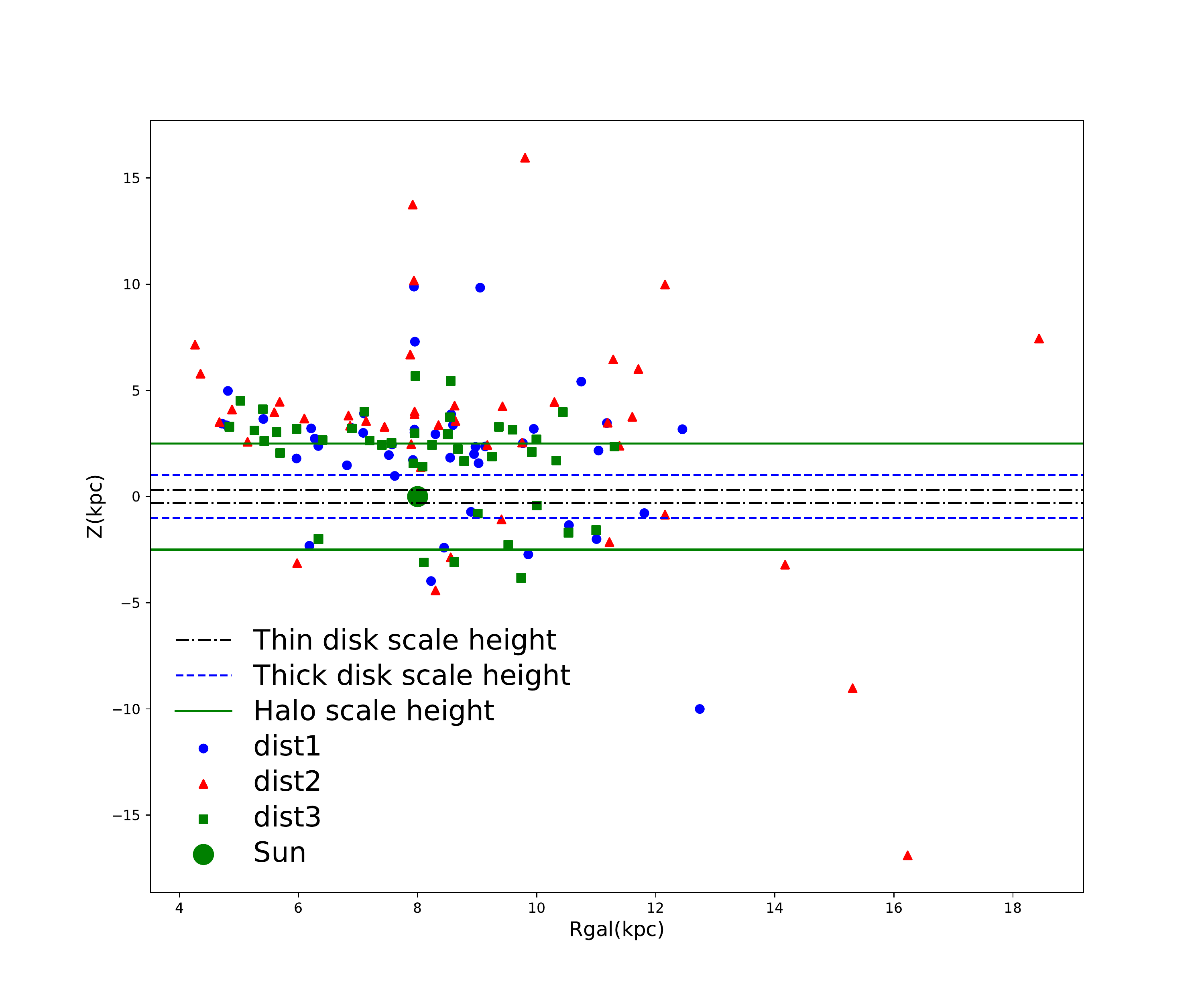} 
\caption{Spatial distribution of CH-normal stars in the meridional Galactic coordinate. Blue dots, red triangles, and green squares indicate the spatial distribution of stars using dist1, dist2, and dist3, respectively (see text). The scale heights of thin disk, thick disk, halo are indicated by black dot-dashed lines, blue dashed lines, green solid lines, respectively. The solar position is labelled by a green filled circle.
}\label{fig:disrz}
\end{figure} 

 Since CH stars may be related to different astrophysical process compared to CH-normal stars, and to avoid confusion in distinguishing two kinds of stars when calculating distances and orbits, we study the kinematics of only CH-normal stars.
To visualize the spatial distribution of our CH-normal stars, we estimate stellar distances using three methods: (1) Bayesian spectro-photometric distances with no assumptions about the underlying populations \citep{Carlin2015}, hereafter dist1; (2) Bayesian spectro-photometric distances with flexible Galactic stellar-population priors \citep{Queiroz2018}, hereafter dist2; (3) Bayesian Gaia DR2 parallax-based distances \citep{BailerJones2018}, hereafter dist3. The basic idea of deriving a Bayesian spectro-photometric distance for a star is to first estimate stellar absolute magnitude in one band from fitting its stellar parameters (T$_{\rm eff}$, $\log$ g, and [Fe/H]) to a given isochrone; then, the distance is given by the distance modulus: $m-M$, where $m$ and $M$ are apparent and absolute magnitudes, respectively. However, detailed treatments can be different between methods. For example, dist1 assumes no prior about the underlying populations, either due to predictions of the luminosity function from stellar evolution modeling, or from Galactic models of stellar populations along each line of sight, while dist2 assumes flexible Galactic stellar-population priors.
Figure \ref{fig:histdist} shows that dist2 method predicts the most extended distance distribution, e.g., the maximum distance of our CH-normal stars is up to about 19 kpc; while dist3 method predicts the most compact distance distribution, e.g., the maximum distance is less than 7.5 kpc; dist1 distances are between these two. Because dist3 is based on parallax from Gaia DR2, the uncertainties increase significantly for stars beyond $\sim 5$ kpc. The parallax-based distances of stars further than that are expected to be dominated by the prior assumed in \citet{BailerJones2018}. 

Using the distances derived by three different methods, we further calculate their positions in the meridional Galactic coordinate.
The results are shown in Figure \ref{fig:disrz}. The spatial distributions using distances from different methods are generally similar for stars within 5 kpc from the Sun, but the differences increase significantly when stars are located further away from the Sun. This is exactly what we expect from the distance distribution (Figure \ref{fig:histdist}). To put our CH-normal stars into the context of our Galaxy, we assume scale heights of the thin disk, the thick disk, and the halo as 0.3 kpc, 1 kpc, and 2.5 kpc, respectively \citep{Sparkebook}. Figure \ref{fig:disrz} indicates no star belongs to the thin disk. Only a few stars have Galactic Z comparable to the thick disk scale height. Most of the stars have Galactic Z comparable to the halo scale height. In other words, our CH-normal stars are mostly halo stars. Note that our CH-normal stars are asymmetric in the Z-direction, because Most of the LAMOST targets are located in the northern Galactic hemisphere. 

\section{Tracking Orbits of CH-normal  Stars}
\label{sect:orb}

\begin{figure*}
\centering
\includegraphics [width=0.9\textwidth]{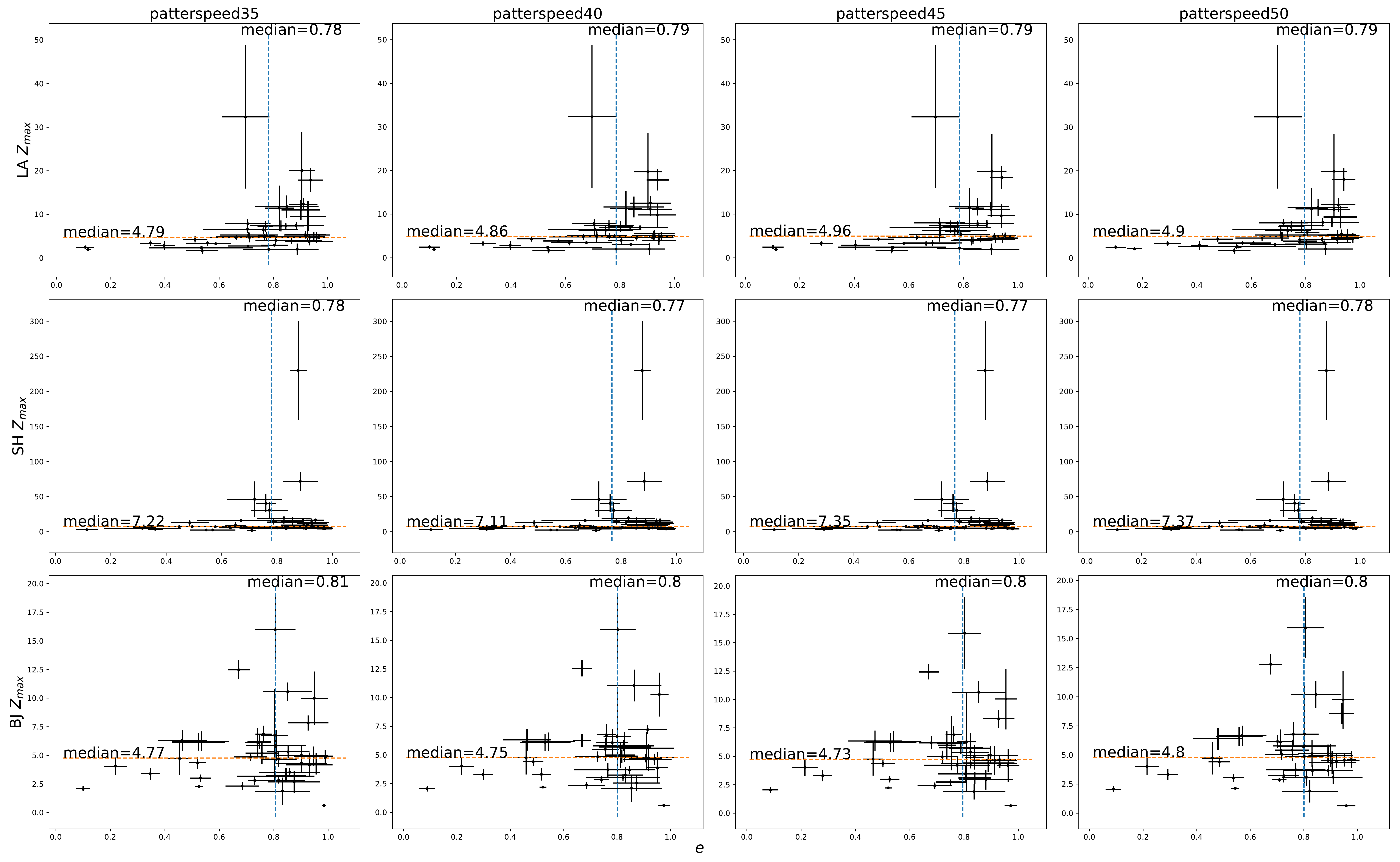} 
\caption{Mean $e$ (eccentricity) versus mean $Z_{\rm max}$ (maximum distance from the Galactic plane) of $10^4$ MC simulation orbits for our CH-normal stars. The 16th and 84th percentile values of each star are indicated by errorbars. Each column of panels assume the same pattern speed (from left to right, 35, 40, 45, and 50 km s$^{-1}$ kpc$^{-1}$), while each row of panels use the distances derived from the same method (from top to bottom, dist1, dist2, and dist3, or LA, SH, and BJ for short). The medians over mean $e$ and mean $Z_{\rm max}$ of our sample stars are shown on the top and left of each panel, respectively.
}\label{fig:orb1}
\end{figure*} 

\begin{figure*}
\centering
\includegraphics [width=0.9\textwidth]{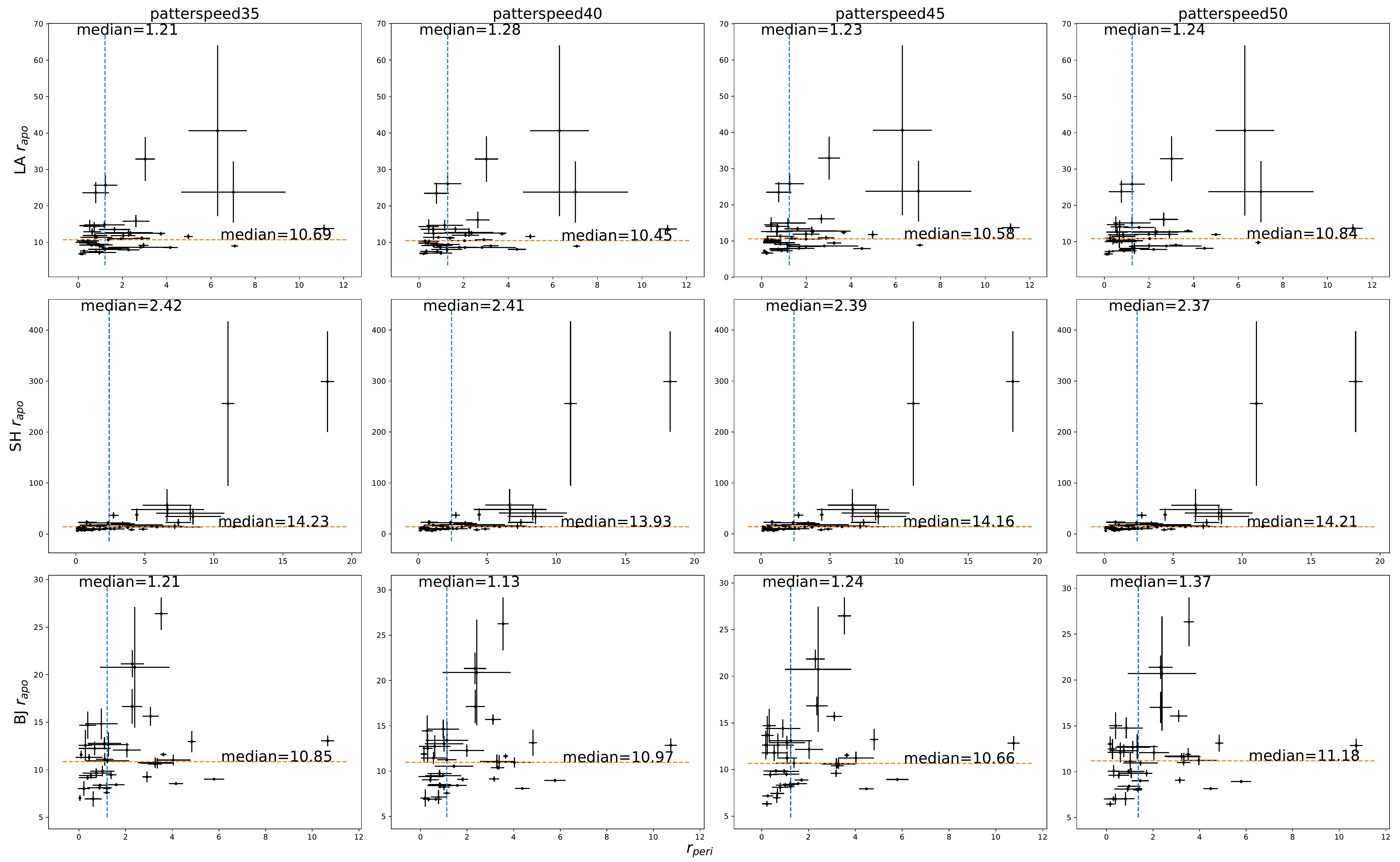} 
\caption{Mean $r_{\rm peri}$ versus mean $r_{\rm apo}$ (perigalactic and apogalactic radii) of $10^4$ MC simulation orbits for our CH-normal stars. The 16th and 84th percentile values of each star are indicated by errorbars. Each column of panels assume the same pattern speed (from left to right, 35, 40, 45, and 50 km s$^{-1}$ kpc$^{-1}$), while each row of panels use the distances derived from the same method (from top to bottom, dist1, dist2, and dist3, or LA, SH, and BJ for short). The medians over mean $r_{\rm peri}$ and mean $r_{\rm apo}$ of our sample stars are shown on the top and right of each panel, respectively.
}\label{fig:orb2}
\end{figure*}

\begin{figure}
\centering
\includegraphics [width=0.5\textwidth]{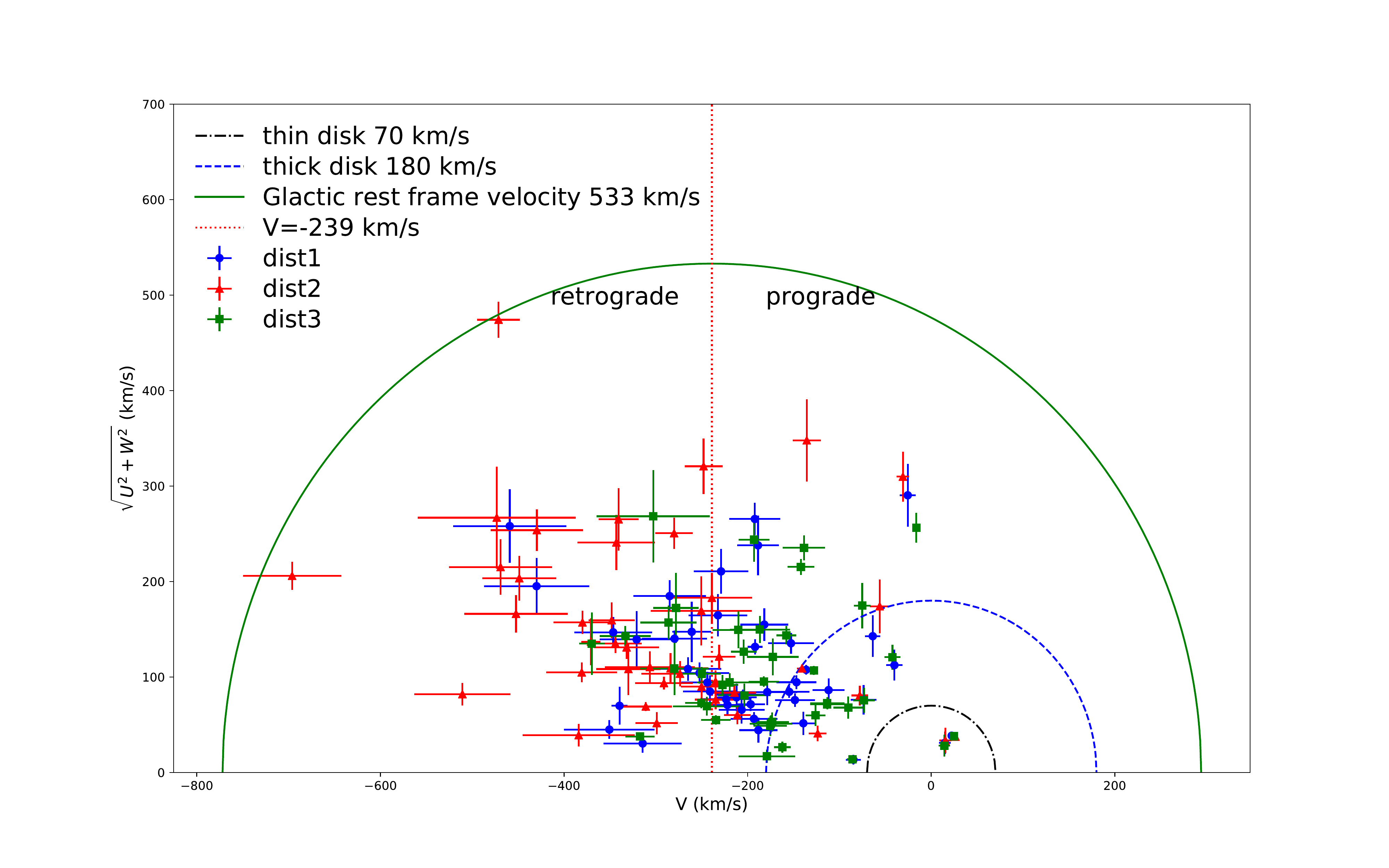} 
\caption{Toomre diagram. The thin disk region is indicated by a constant velocity of 70 km/s (black dot-dashed curve), and the thick disk region is indicated by a constant velocity of 180 km/s (blue dashed curve). A constant galactic rest frame velocity of 533 km/s is shown as green curve. Finally, $\rm V=-239$ km/s is used to distinguish prograde and retrograde orbits (red dotted line).
}\label{fig:too}
\end{figure}

In this work, we simulate the orbits of our CH-normal stars using a state-of-the-art orbital integration model in a (as far as possible) realistic Galactic gravitational potential. To do this, we employ the novel dynamical model called GravPot16 (Fern\'andez-Trincado et al., in prep.), which has been adopted in a score of papers \citep{Fernandez-Trincado2016, Fernandez-Trincado2017b,  Tang2018, SU2018}. The Galactic model of GravPot16 is briefly summarized in the Appendix of \citet{Tang2018}. A long list of studies in the literature has presented different ranges for the bar pattern speeds \citep{Portail_2017, Monari_2017a, Monari_2017b, Fernandez-Trincado_2017b}. For our computations, 
we assume four pattern speeds, $\Omega_{\rm B}=35, 40, 45, 50$ km s$^{-1}$ kpc$^{-1}$. 
To obtain robust estimates of the orbits of our CH-normal stars, we employ recent high-quality data as input parameters for GravPot16. 
We use the radial velocities and absolute proper motions from the latest Gaia DR2 \citep{Brown2018, Katz2018}. The typical uncertainty of radial velocities of our CH-normal stars is 0.8 km/s, while the typical uncertainty of absolute proper motions is 0.05 mas/yr. Besides, we assume distances derived from three different methods as we mention above (dist1, dist2, and dist3).

After combining our Milky Way potential model with measurements of radial velocity, absolute proper motion, distance, and sky position for the CH-normal stars, we run $N_{\rm total}=10^4$ orbit simulations for each of the CH-normal stars, taking into account the uncertainties in the input data considering $1\sigma$ variations in a Gaussian Monte Carlo (MC) approach. For each generated set of parameters, the orbit is computed backward in time, up to 2.5 Gyr. The major assumptions and limitations in our computations are also discussed in \citet{Tang2018} and \citet{SU2018}. From the integrated set of orbits, we compute (1) $r_{\rm peri}$,  the perigalactic radius, (2) $r_{\rm apo}$, the apogalactic radius, (3) the orbital eccentricity, defined as $e=(r_{\rm apo} - r_{\rm peri})/(r_{\rm apo} + r_{\rm peri})$, (4) the maximum vertical amplitude $Z_{\rm max }$.

Figures \ref{fig:orb1} and \ref{fig:orb2} show the $r_{\rm peri}$, $r_{\rm apo}$, $e$, and $Z_{\rm max }$ of our sample stars. Each error bar indicates the mean of a given parameter for each star among the $10^4$ MC orbit realizations with uncertainty range given by the 16th and 84th percentile values. Each column of panels assume the same pattern speed (from left to right, 35, 40, 45, and 50 km s$^{-1}$ kpc$^{-1}$), while each row of panels use the distances derived from the same method (from top to bottom, dist1, dist2, and dist3, or LA, SH, and BJ for short).  
We find that (1) pattern speed has negligible impact on the four parameters that we investigate; (2) distance has negligible impact on $e$, and small impact on $r_{\rm peri}$; (3) distance has substantial impact on $r_{\rm apo}$ and $Z_{\rm max }$. For example, $Z_{\rm max }$ derived using dist2 (median$\sim$7.2 kpc) are systematically greater than that of dist1  (median$\sim$4.9 kpc), and that of dist3 (mean$\sim$4.7 kpc). This is consistent with their distance distributions (Figure \ref{fig:histdist}). Independent of which distance is assumed, most of our CH-normal stars lie on highly eccentric ($e \sim 0.8$) orbits with large $r_{\rm apo}$ and $Z_{\rm max }$, which is consistent with halo star kinematics.

We further employ the Toomre diagram to identify the origin of our CH-normal stars. Figure \ref{fig:too} shows that Galactic velocity distributions of stars strongly depend on the adopted stellar distances. In general, a few stars belong to the Galactic thick/thin disks, but most stars belong to the Galactic halo. This is similar to the conclusion that we draw from the spatial distribution in the meridional Galactic coordinate ($\S$ \ref{sect:sp}). We also find that a substantial fraction of stars lie in retrograde orbits --- dist2 indicate more stars in retrograde orbits, and more extended distribution of V; while less retrograde orbiting stars in the case of dist3. Accordingly, dist1, dist2, and dist3 indicate 33\%, 67\%, and 25\% stars lying in retrograde orbits, respectively.
The reader is reminded that dist3 is based on parallax of Gaia DR2, which is less certain for stars further than 5 kpc from our Sun.

\section{Discussion}
\label{sect:dis}

\subsection{Compared with Literature}

As we mentioned in Section \ref{sect:intro}, several literature studies have search for N-rich stars or chemically peculiar stars. After matching our CN-strong stars with the sample of M11, we found no common star. Because \citet{Martell2016, Fernandez-Trincado2016}  and \citet{FT2017} used APOGEE data, their stars should be available in APOGEE DR14. However, the 8 common CN-strong stars that we identified between LAMOST and APOGEE are not found in the aforementioned studies. 
S17 targeted inner Galaxy stars with $-20^{\circ}<l<20^{\circ}$, $|b|<16^{\circ}$. This part of the sky is not included in our work, thus no common star is found. We further confirm that there is no overlap between our sample and the chemically peculiar stars found in \citet{Carretta2010b, Ramirez2012, Lind2015}. To summarize, we find no common star between those in published studies and our sample stars. Therefore, we report for the first time 44 CN-strong CH-normal stars using LAMOST DR3, along with 35 stars identified as CN-strong CH-strong (five of them have been identified as carbon stars by \citealt{LiYB2018}).

M11 share a few similarities with our work, but our work has several major advantages: (1) three times larger metal-poor field star sample size, thanks to the superior data-collecting capability of LAMOST; (2) using both CN3839 and CN4142 to select CN-strong stars, followed by visual inspection; separating CH stars from CH-normal stars, and discussing the different chemical pattern of these CH stars; (3) including HK$^{\prime}$ in the analysis; (4) combining the chemical abundances derived by high resolution APOGEE data and our data-driven software to understand their origin, especially with C-N parameter space; (5) state-of-the-art orbital integration model with a realistic Galactic gravitational potential. We further note that \citet{Carollo2013} suggested that the stars in M11 are mostly in prograde orbits, while we find a substantial fraction of stars (25\%-67\% depending on the adopted distances) lie in the retrograde orbits. Finding no common stars between M11 and our CH-normal star sample is also unexpected, since two works use similar experimental logic and search the northern sky. We checked SDSS DR10 database, and found that none of the 44 CH-normal stars has existing spectrum. In fact, the sky coverage of LAMOST survey is much larger than that of SEGUE and SEGUE-2.

S17 found a large population of N-rich stars in the inner Galaxy with APOGEE data. A few C-enhanced stars ([C/Fe$]>+0.15$) are also seen in their work, and they excluded these group of stars to concentrate on the N-rich stars with normal C abundances. We also found a non-negligible number of CH stars in our study, and these stars show distinct chemistry compared to other field stars.

The control sample of bulge field stars presented in S17 shows an N-C anti-correlation between $-0.5<[$C/Fe$]<-0.1$ (Figure 2 of their paper), while their N-rich stars are above this sequence. S17 showed that this sequence follows the nitrogen lower envelope defined by the giants from several halo GCs. Thus, they suggested this sequence is caused by the extra mixing process during the RGB phase, while their N-rich stars cannot be produced by only extra mixing due to their high N abundances.
We see a similar configuration in the left panel of Figure \ref{fig:abun}. A lower N-C sequence is outlined by normal metal-poor field stars, where an anti-correlation is seen. The CH-normal stars are clearly above this sequence. On the other side, we obtain chemical abundance information from low-resolution LAMOST data with our data-driven software, SLAM. A similar configuration is seen in the N-C parameter space (Figure \ref{fig:lcn}).  We further showed that brighter normal metal-poor field stars tend to have higher N abundances, which is consistent with extra mixing theories \citep{Charbonnel1994,  Boothroyd1999, Charbonnel2010}.
Thus the C and N abundances support that these CH-normal stars have different enrichment history than normal halo field stars, because extra mixing process cannot be fully responsible for the high N abundances. 

\citet{FT2017} found a group of N, Al-enhanced, but Mg-depleted stars. The stars in their metal-poor sample ([Fe/H$]<-1.0$) show similar chemical pattern as some SG Galactic GC stars. In our common stars with APOGEE, we did not find Mg-depleted stars, nor high Al abundance ([Al/Fe$]>+0.5$) stars as found in \citet{FT2017}. From the chemical point of view, our CN-strong stars may not be the same kind of stars as their metal-poor sample.

\subsection{Are CH-normal Stars Dissolved GC Stars?}

If we assume that CH-normal stars are SG stars from dissolved GCs, we can estimate the contribution of dissolved GC members to the field. In this scenario, the FG stars are buried inside the halo field stars, and they cannot be easily distinguished. Here we follow the minimal and maximal scenarios suggested by S17. The minimal scenario assumes the present day FG/SG ratio is about 1/2, and the maximal scenario assumes an extreme condition, where FG/SG is about 9/1. According to our study, we find 44 CH-normal stars out of 7723 field stars, which accounts for 0.6\% of the field stars. This ratio is smaller than what was found in other studies, e.g., M11, \citet{Ramirez2012} (2.5-3\%). We note that our sample selection requires that both CN3839 and CN4142 spectral indices are significantly (2$\sigma$) higher than other normal metal-poor field stars, which is more stringent compared to other studies.
Under the minimal scenario, we expect to find a total of 66 dissolved GC member stars, which account for  $\sim0.9$\% of the field stars. This number is slightly smaller than the value found by S17, about 1.7\%. Next, under the maximal scenario, early GCs are expected to lose $\sim90$\% of their mass before the formation of SG stars \citep{Bastian2015}. We expect to find a total of 440 stars as dissolved GC members, which account for  $\sim6$\% of the field stars. However, the maximal scenario was basically ruled out in S17 due to (1) different MDFs (metallicity distribution functions) between their N-rich stars and the rest of the inner Galaxy population; and (2) that the dissolved GC members outweigh the GC system by a factor of 80.

\begin{figure}
\centering
\includegraphics [width=0.50\textwidth]{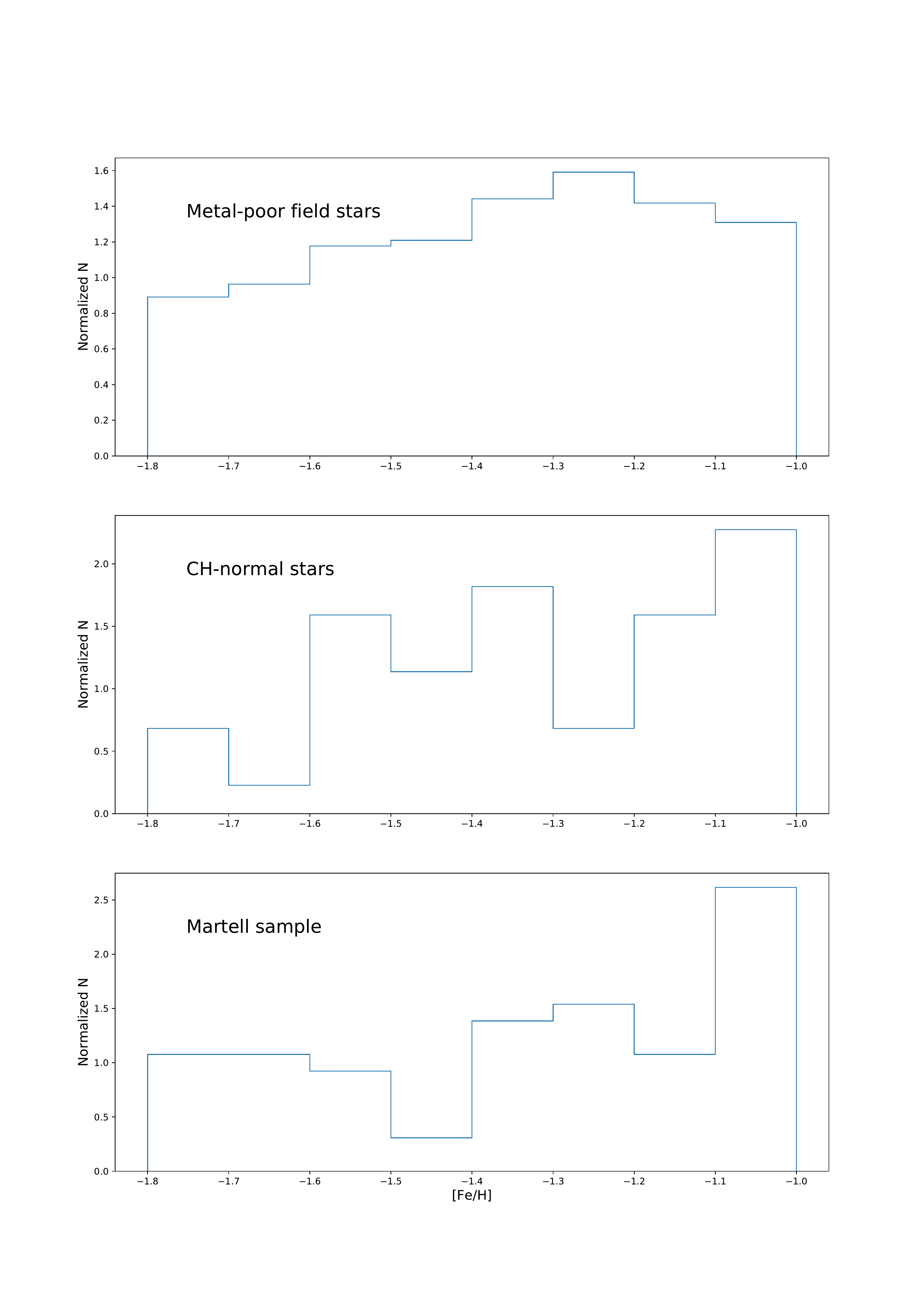} 
\caption{Metallicity distribtution of normal metal-poor field stars, CH-normal stars, and M11 stars.}\label{fig:mdf}
\end{figure}

The maximal scenario needs revision in our work, as we are dealing with metal-poor field stars, not inner Galaxy field stars, like S17. First, we plot the MDFs of the CH-normal stars and of the metal-poor field stars in Figure \ref{fig:mdf}. The two distributions both have more stars towards higher metallicity. The metal-poor field stars seem to peak around [Fe/H$]=-1.2$, while the peak of CH-normal stars could be around [Fe/H$]=-1.1$, but it is not totally clear due to low statistics and limited metallicity range. At least from what we have, the two MDFs are not strongly different. 
Next, if we adopt the Galactic GC system mass from \citet{Kruijssen2009} ($\sim2.8 \times10^7 M_{\odot}$), and the luminous Galactic halo mass from \citet{BlandHawthorn2016} ($\sim 4-7 \times10^8 M_{\odot}$), then the mass of dissolved GC members ($\sim 1-6$\% of the halo field stars) is comparable to the current GC system (the ratio is between $0.14-1.5$). 

 Interestingly, the MDF of the N-rich stars found in the inner Galaxy peaks around [Fe/H$]=-1.2  \mbox{---} -1.0$, which shows no obvious contradiction with the MDFs of the CH-normal stars and of the metal-poor field stars in our work. Next, the number ratio of the N-rich stars over the inner-Galaxy field stars and the number ratio of our CH-normal stars over the halo field stars are similar. Better agreement may be reached if we consider that our selection method CN-strong stars is more stringent. Furthermore, our two CH-normal stars (common stars with APOGEE DR14) have compatible [C/Fe], [N/Fe], [Mg/Fe], and [Al/Fe] with the S17 N-rich stars.  Is it possible that the N-rich stars found by Schiavon et al. are the same stellar population as the CH-normal stars that we find? May be S17 N-rich stars are halo stars streaming towards the Galactic center? More observational evidence (e.g., kinematic analysis of the S17 sample stars, high-resolution spectra of our CH-normal stars) is needed to verify this statement.
What about the M11 stars? We find that they share similar MDF as our CH-normal stars (Figure \ref{fig:mdf}), though their kinematic signatures may be different \citep{Carollo2013}. It would be interesting to run MC orbit simulations over all literature samples to eliminate the model-dependence of different studies. 
If these stars are found over the whole Galaxy, it implies their formation may be quite early, and now they have reached dynamic equilibrium. 

\subsection{Are CH-normal Stars Accreted Halo Stars?}

Recent studies have shown that the Galactic halo stars ([Fe/H$]<-0.9$) may be separated into two populations with distinct chemical and kinematic signatures \citep{Nissen2010, Ramirez2012, Hayes2018}. Two populations of stars are called ``high-$\alpha$ population'' and ``low-$\alpha$ population'' based on their $\alpha$-element abundances. The ``low-$\alpha$'' stars tend to have lower O, Mg, Si, Na, Ni, Cu and Zn and higher Eu abundances than those of the thick disk stars and ``high-$\alpha$'' stars at similar metallicity. Kinematically, the ``low-$\alpha$'' stars have little to no net rotation \citep{Hayes2018}, and some of them may also lie in retrograde orbits \citep{Nissen2010}. These kinematics suggest that ``low-$\alpha$'' stars are accreted from nearby dwarf galaxies, and thus, they are also called accreted halo stars.  In this work, we find that a substantial fraction of our CH-normal stars are retrograding (25\% to 67\% depending on the adopted distances), which is a typical signature of accreted halo stars. Chemically, we find that our CH-normal stars are N-enriched compared to the bulk metal-poor halo field stars. Further investigation of their chemical patterns need to wait for high-resolution spectra for the whole CH-normal star sample. 

Some GCs are suggested to be accreted from nearby dwarf galaxies, e.g. Sagittarius GCs \citep{Law2010} and Canis Major GCs \citep{Martin2004,Penarrubia2005}. The latter ones may also be referred as ``Sausage'' GCs \citep{Myeong2018}. \citet{Kruijssen2018} further proposed $10-20$ GCs were brought in the Milky Way by a massive satellite, Kraken, $6-9$ Gyr ago.  Particularly, some of the Sagittarius GCs and ``Sausage'' GCs are in fact iron-complex \citep{Dacosta2015}. For example, the iron-complex Sagittarius GC, M 54. Therefore, we speculate some of our CH-normal stars could be stars belonging to accreted GCs from a nearby dwarf galaxy and later dissolved into the field.

\section{Conclusion}
\label{sect:con}

We have identified 79 CN-strong metal poor ([Fe/H$]<-1$) field stars from LAMOST DR3 using CN spectral features around 3839 \AA~and 4142 \AA. The sample was further separated into CH-normal (44) and CH (35) stars based on their CH spectral features around 4300 \AA. This CH-normal star sample was identified for the first time in our study. 
We checked the high resolution chemical abundances from APOGEE for eight common stars, and used our data-driven software, SLAM, to derive C and N abundances from LAMOST spectra for stars with high S/N. We found that CH stars, CH-normal stars, and normal metal-poor field stars have different chemical patterns, especially in the N-C parameter space. 

We adopted distances derived from three methods for CH-normal stars, and integrated their orbits using GravPot16 with a Monte Carlo approach. Their orbital parameters and spatial distribution indicate that CH-normal stars are mostly halo stars. The Toomre diagram suggests that a substantial fraction (25\% to 67\%, depending on the adopted distance) of them are retrograding. 
The chemistry and kinematics of CH-normal stars imply that they may be GC-dissolved stars, or accreted halo stars, or both. In the future, we will increase our sample size by applying our technique to later LAMOST data releases.  High spectral resolution follow-up observations of CH-normal stars are also planned.

\section{acknowledgments}

We thank Anna Barbara Queiroz and Friedrich Anders for computing the star distances. We thank Sarah Martell and A-Li Luo for helpful discussions. 
We thank the anonymous referee for insightful comments. 
BT acknowledges support from the one-hundred-talent project of Sun Yat-Sen University.
JGFT and DG gratefully acknowledge support from the Chilean BASAL Centro de Excelencia en Astrof\'{i}sica y Tecnolog\'{i}as Afines (CATA) grant PFB-06/2007. DG also acknowledges financial support from the Direcci\'{o}n de Investigaci\'{o}n y Desarrollo de la Universidad de La Serena through the Programa de Incentivo a la Investigaci\'{o}n de Acad\'{e}micos (PIA-DIDULS).

Guoshoujing Telescope (the Large Sky Area Multi-Object Fiber Spectroscopic Telescope LAMOST) is a National Major Scientific Project built by the Chinese Academy of Sciences. Funding for the project has been provided by the National Development and Reform Commission. LAMOST is operated and managed by the National Astronomical Observatories, Chinese Academy of Sciences.
Funding for the GravPot16 software has been provided by the Centre national d'etudes spatiale (CNES) through grant 0101973 and UTINAM Institute of the Universit\'e de Franche-Comt\'e, supported by the R\'egion de Franche-Comt\'e and Institut des Sciences de l'Univers (INSU).
Monte Carlo simulations have been executed on computers from the Utinam Institute of the Universit\'e de Franche-Comt\'e, supported by the R\'egion de Franche-Comt\'e and Institut des Sciences de l'Univers (INSU).

\bibliographystyle{aasjournal}
\bibliography{gc,survey,cnchcor}

\label{lastpage}

\end{document}